\documentclass[runningheads]{llncs}

\usepackage[T1]{fontenc}
\usepackage{graphicx}
\usepackage{amssymb}
\usepackage{hyperref}
\usepackage{dsfont}
\usepackage{epstopdf}

\usepackage{floatrow}
\newfloatcommand{capbtabbox}{table}[][\FBwidth]

\usepackage{subcaption} 

\usepackage{amsmath}
\usepackage{multirow}
\usepackage{soul}
\usepackage{booktabs}
\usepackage{bbding}
\usepackage{makecell}
\usepackage{stfloats}
\usepackage{xcolor}
\usepackage{makecell}
\usepackage{xspace}
\usepackage{url}
\usepackage{wrapfig}
\usepackage{multirow}
\usepackage{makecell}
\usepackage{booktabs, colortbl}
\usepackage{longtable}
\usepackage{hhline}
\usepackage{algpseudocode}
\usepackage{listings}
\usepackage[ruled]{algorithm2e}
\definecolor{myblue}{HTML}{0072C6}
\definecolor{myyellow}{HTML}{FFFADF}
\definecolor{myred}{HTML}{FF0000}
\usepackage[misc]{ifsym} 
\newcommand{\para}[1]{\vspace{.05in}\noindent\textbf{#1}}

\begin{document}
\title{Cross-conditioned Diffusion Model for Medical Image to Image Translation}


\titlerunning{Cross-conditioned Diffusion Model}

\author{Zhaohu Xing\inst{1} \and
Sicheng Yang\inst{2} \and
Sixiang Chen \inst{1} \and 
Tian Ye\inst{1} \and 
Yijun Yang\inst{1} \and
Jing Qin\inst{3} \and
Lei Zhu\inst{1,4} \textsuperscript{(\Letter)}
}


\institute{The Hong Kong University of Science and Technology (Guangzhou) \and
{Xi’an Jiaotong University} \and
{The Hong Kong Polytechnic University} \and
The Hong Kong University of Science and Technology \\
\email{leizhu@ust.hk} 
}

\maketitle              
\begin{abstract}
Multi-modal magnetic resonance imaging (MRI) provides rich, complementary information for analyzing diseases. 
However, the practical challenges of acquiring multiple MRI modalities, such as cost, scan time, and safety considerations, often result in incomplete datasets. This affects both the quality of diagnosis and the performance of deep learning models trained on such data. 
Recent advancements in generative adversarial networks (GANs) and denoising diffusion models have shown promise in natural and medical image-to-image translation tasks. However, the complexity of training GANs and the computational expense associated with diffusion models hinder their development and application in this task. 
To address these issues, we introduce a Cross-conditioned Diffusion Model (CDM) for medical image-to-image translation. 
The core idea of CDM is to use the distribution of target modalities as guidance to improve synthesis quality while achieving higher generation efficiency compared to conventional diffusion models.
First, we propose a Modality-specific Representation Model (MRM) to model the distribution of target modalities. Then, we design a Modality-decoupled Diffusion Network (MDN) to efficiently and effectively learn the distribution from MRM. Finally, a Cross-conditioned UNet (C-UNet) with a Condition Embedding module is designed to synthesize the target modalities with the source modalities as input and the target distribution for guidance. Extensive experiments conducted on the BraTS2023 and UPenn-GBM benchmark datasets demonstrate the superiority of our method.

\keywords{Cross-conditioned model  \and Diffusion model \and Multi-modal MRI \and Medical image to image translation.}
\end{abstract}

\vspace{-1mm}
\section{Introduction}
\vspace{-1mm}
Multi-modal magnetic resonance imaging (MRI) is crucial for the comprehensive analysis and diagnosis of diseases and is routinely used in clinical settings~\cite{wang2024advancing,wang2023video,zhao2023uncertainty,liu2023contrastive,liu2019probabilistic,xing2024segmamba,xing2024hybrid,xing2023diff}. They provide rich, complementary information for analyzing brain tumors.
Specifically, for gliomas, the commonly used MRI sequences including T1-weighted (T1), post-contrast T1-weighted (T1Gd), T2-weighted (T2), and T2 Fluid Attenuation Inversion Recovery (T2-FLAIR) images~\cite{xing2022nestedformer,luo2023segrap2023}. Each sequence plays a varying role in distinguishing between the tumor, peritumoral edema, and the tumor core.
However, obtaining multiple modalities in clinical settings can be challenging due to factors such as scan costs, limited scan time, and safety considerations. 
Consequently, the absence of certain crucial modalities can have a detrimental impact on the quality of diagnosis and treatment. 
Furthermore, deep learning models based on multi-modal MRIs also suffer from decreasing performance when crucial modalities are missing in training data.

Generative adversarial networks (GANs)~\cite{isola2017image,zhu2017unpaired,fu2019geometry,liu2020psi} have been extensively explored for natural image-to-image translation.
However, these methods are difficult to apply directly to medical imaging due to the domain gap between natural and medical images.
To address this issue, RegGAN~\cite{kong2021breaking} uses an additional registration network to fit the misaligned noise distribution.
ResViT~\cite{dalmaz2022resvit,liu2023scotch} proposes a transformer-based central bottleneck module designed to distill task-critical information while preserving both global and local information within high-dimensional medical images.
Although there has been significant development in these methods, the training process of GANs is not stable.

%
Recently, denoising diffusion models~\cite{ho2020denoising,yang2024vivim,song2020improved,hu2024diffusion,nan2023hunting,ye2024learning}, which are capable of offering better details, have shown significant success in various generative tasks. 
%
%
Dhariwal~\cite{dhariwal2021diffusion} et al. propose the first diffusion model with conditional input and achieves better performance compared to GANs. 
However, diffusion models introduce additional computational costs since they must sample multiple times during inference.
RCG~\cite{li2023self} introduces the concept of self-conditioned image synthesis for the first time and outperforms conventional diffusion models in terms of accuracy and efficiency.
However, RCG cannot be directly applied to image-to-image translation tasks due to its self-conditioning mechanism.
Moreover, the encoder and decoder pre-trained on natural images in RCG are not well-suited for medical images.

\begin{figure}[t]
\includegraphics[width=0.9\columnwidth]{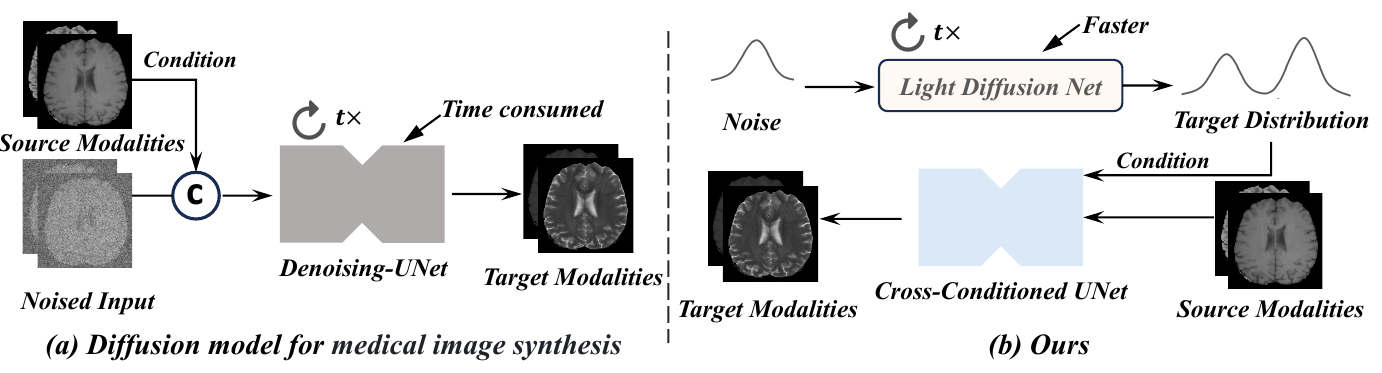}
\caption{
Comparison between the conventional Diffusion model (a) and our method (b). Our method replaces the time-consuming denoising UNet with a light Diffusion network, which achieves higher efficiency. 
} 
\label{fig:intro}
\end{figure}
In this paper, we propose a Cross-conditioned Diffusion Model for medical image-to-image translation, named CDM.
Instead of directly sampling the target modalities in the image domain like the conventional diffusion model, CDM first samples the distribution of target modalities in latent variable space, and then this distribution is used as a condition to generate the target modalities in the image domain.
First, we design a Modality-specific Representation Model (MRM) to learn the distribution of target modalities.
Subsequently, as illustrated in Fig. \ref{fig:intro}, we replace the time-consuming denoising UNet with a light Diffusion network, called Modality-decoupled Diffusion Network (MDN), which achieves higher efficiency in both training and inference and model the target distribution from MRM.
Finally, we propose a Cross-conditioned UNet (C-UNet) with a Condition Embedding module to receive the source modalities and distribution sampled by MDN as input to generate the target modalities.
Extensive experiments on BraTS2023~\cite{menze2014multimodal,bakas2017advancing,kazerooni2023brain} and UPenn-GBM~\cite{bakas2022university,li2024tp} datasets demonstrate the superiority of our proposed CDM.

\vspace{-1mm}
\section{Method}

Our CDM primarily consists of three components: 1) the modality-specific representation model, which learns the distribution of target modalities; 2) the modality-decoupled diffusion network, designed for improved feature representation and efficiency; and 3) the cross-conditioned UNet model, which generates target modalities from source modalities and sampled target distribution.
%
%

\begin{figure}[t]
\includegraphics[width=0.95\columnwidth]{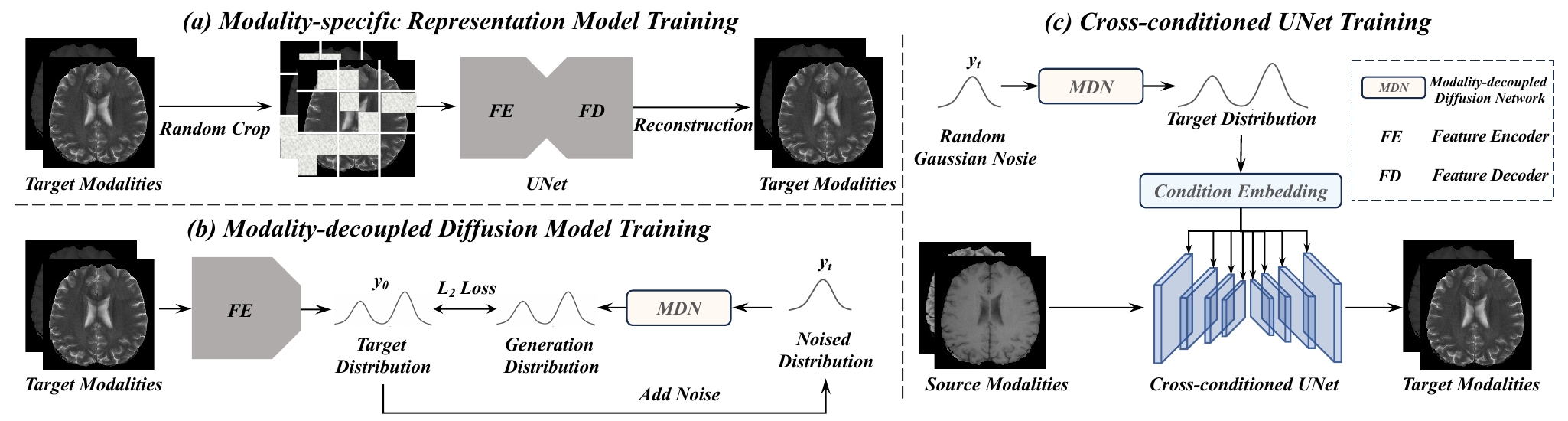}
\caption{
An overview of the proposed Cross-conditioned Diffusion Model (CDM). First, we introduce the Modality-specific Representation Model (a) to learn the distribution of target modalities. Then, the Modality-decoupled Diffusion Network (b) is employed to learn the target distribution. Finally, the Cross-conditioned UNet (c) incorporates the source modalities and samples the target distribution as guidance to generate the target modalities.
} 
\label{fig:method}
\end{figure}

\vspace{-1mm}
\subsection{Representation Learning for Target Modalities}

\para{Modality-specific Representation Model (MRM) Training}
In RCG, both the feature encoder and decoder are pre-trained on natural images, which results in reduced performance when processing medical images. We design a modality-specific representation model consisting of a feature encoder FE and decoder FD.
Similar to SimMIM~\cite{xie2022simmim}, as shown in Fig. \ref{fig:method} (a), we randomly mask some patches in each target modality separately and concatenate them at the channel dimension as input.
Then, the MRM learns to restore the original target modalities, supervised by the $L_2$ loss function.
\begin{equation}
\vspace{-2mm}
    \mathcal{L}_{\mathrm{MRM}} = \frac{1}{|{R}|}\sum_{r \in {R}} || p_{r} - \hat{p}_{r}||_2,
\end{equation}
where $R$ denotes masked patches in target modalities, $|R|$ denotes the number of masked patches, $p_r$ and $\hat{p_r}$ represent the prediction values and input values.

\begin{figure}[t]
\includegraphics[width=0.95\columnwidth]{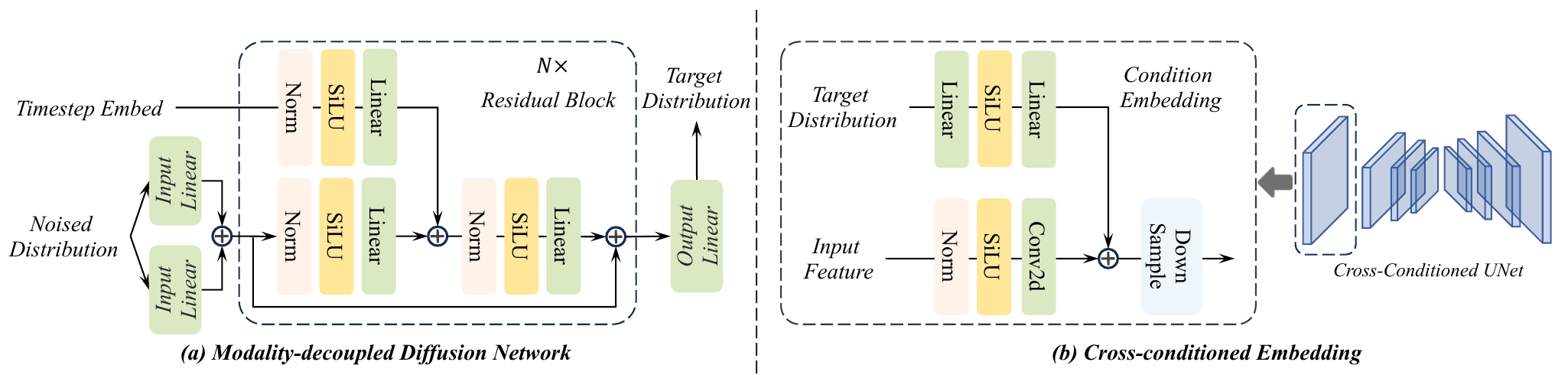}
\caption{
An overview of Modality-decoupled Diffusion Network (a) and Cross-conditioned Emebedding (b).
} 
\label{fig:ldn}
\vspace{-3mm}
\end{figure}

\para{Modality-decoupled Diffusion Network (MDN) Training}
The core of cross-conditioned image generation lies in using the target distribution sampled by the diffusion model to guide the pixel generation process for target modalities.
To achieve this, we adopt a light modality-decoupled diffusion network to efficiently sample the target distribution.
As shown in Fig. \ref{fig:ldn} (a), the MDN first employs two separate linear layers to decouple the noised target distribution $y_t$. 
Then, multiple residual blocks are utilized to eliminate the noise and generate target distribution $y_0$.
Each residual block consists of an input layer, a timestep embedding layer, and an output layer, where each layer comprises a LayerNorm~\cite{xu2019understanding}, a SiLU~\cite{paul2022sinlu}, and a linear layer.
The MDN follows Denoising Diffusion Implicit Models (DDIM)~\cite{song2020denoising} for training and inference. As shown in Fig. \ref{fig:method} (b), during training, the target distribution $y_0$ from feature encoder FE is mixed with Gaussian noise $\epsilon$ over $t \in \{0,1,...,T\}$ steps. 
\begin{equation}
    q\left({y}_t \mid {y}_0\right)=\mathcal{N}\left({y}_t ; \sqrt{\bar{\alpha}_t} {y}_0,\left(1-\bar{\alpha}_t\right) \epsilon\right),
\end{equation}
where $\bar{\alpha}_t = \prod_{s=0}^t \alpha_s=\prod_{s=0}^t\left(1-\beta_s\right)$  and $\beta_s$
represent the noise schedule~\cite{ho2020denoising}.
%
Then, the MDN learns to restore ${y}_0$ from noised ${y}_t$ supervised by $L_2$ loss function.
During inference, as shown in Fig. \ref{fig:method} (c), the target distribution ${y}_0$ is predicted by the MDN from a normal Gaussian noise ${y_t}$, along with the sample schedule~\cite{song2020denoising}.
\begin{equation}
    p_\theta\left({y}_{0: T} \right)=p\left({y}_T\right) \prod_{t=1}^T p_\theta\left({y}_{t-1} \mid {y}_t\right),
\end{equation}
where $\theta$ denotes the training parameters of the MDN.

\vspace{-1mm}
\subsection{Cross-conditioned UNet (C-UNet)}

To incorporate the generated target distribution into the UNet model, we propose a cross-conditioned UNet, featuring a cross-conditioned embedding module designed to merge the target distribution with the input feature at each scale.
As depicted in Fig. \ref{fig:ldn} (b), for a layer in C-UNet, the input feature from source modalities is fed into a convolution block, which consists of a LayerNorm, a SiLU, and a $1\times 1$ convolution layer, to obtain the feature representation.
Simultaneously, the generated target distribution is processed through an multi-layer perceptron (MLP)~\cite{tolstikhin2021mlp} layer, consisting of two linear layers and a SiLU, to produce the distribution representation.
The feature representation is then combined with the distribution representation for fusion, and a down-sampling layer is utilized to reduce the spatial dimensions of the fused feature.

The final synthesis loss $\mathcal{L_\mathrm{Syn}}$ consisting of mean square error is calculated on the prediction $\hat{Y}$ by the C-UNet and corresponding ground truth $Y$:
\begin{equation}
\vspace{-2mm}
    \mathcal{L_\mathrm{Syn}} = || \hat{Y} - Y||_2.
\end{equation}


\vspace{-2mm}
\section{Experiments}

\subsection{Datasets and Implementation}
\para{BraTS2023 dataset}
The BraTS2023 dataset~\cite{menze2014multimodal,bakas2017advancing,kazerooni2023brain} contains a total of 1,251 3D brain MRI volumes. For each patient, multi-parametric magnetic resonance imaging (mpMRI) scans are available, including the four structural MRI scans: native T1-weighted (T1), post-contrast T1 (T1-Gd), native T2-weighted (T2), and T2 fluid attenuated inversion recovery (T2-FLAIR) scans.

\para{UPenn-GBM dataset}
The UPenn-GBM dataset~\cite{bakas2022university} is composed of 630 patients diagnosed with Glioblastoma Multiforme (GBM). Each volume also includes four modalities (namely T1, T1Gd, T2, and T2-FLAIR).

\noindent For both mentioned datasets, we denote T1Gd as T1c, and T2-FLAIR as T2f. We partition the 3D image along the Z-axis into 2D images. The T1 and T2 modalities are utilized to generate the T1c and T2f modalities.

\para{Implementation details} 
Our model is implemented in PyTorch 2.1.0-cuda12.1. During
training, we resize each image to $256\times 256$ and use a batch size of 12 per GPU for each dataset. 
We employ cross-entropy loss and adopt the Adam optimizer with a learning rate of 1e-4 and a decay rate of 1e-5. 
We run 100 epochs for all datasets. All experiments are conducted on a cloud computing platform with NVIDIA A100 GPUs. For each dataset, we randomly allocate 70\% of the 3D volumes for training, and the remaining 30\% for testing.

\vspace{-2mm}
\subsection{Comparison with SOTA Methods}
We compare our CDM with seven state-of-the-art synthesis methods, including five CNN-based methods (Pix2pix~\cite{isola2017image}, CycleGAN~\cite{zhu2017unpaired}, GcGAN~\cite{fu2019geometry}, CUT~\cite{park2020contrastive}, RegGAN~\cite{kong2021breaking}), one transformer-based method (ResViT~\cite{dalmaz2022resvit}), and one diffusion-based method (conditional diffusion model~\cite{dhariwal2021diffusion}, denoted as 'Diffusion' in all tables).
For a fair comparison, we utilize public implementations of
these methods to retrain their networks, generating their best synthesis results.
The Peak Signal-to-Noise Ratio (PSNR)~\cite{hore2010image}, Structural Similarity Index (SSIM)~\cite{sara2019image}, and Mean Absolute Error (MAE) are adopted for quantitative comparison on the BraTS2023 and UPenn-GBM datasets.

\begin{table*}[!t]
    \centering
    \caption{Quantitative comparison on BraTS2023 dataset} 
    \label{tab:bras2023}
    \renewcommand\arraystretch{1.2}
    \resizebox{0.9\textwidth}{!}{
    \begin{tabular}{c | c c c | c c c | c c c}
    \toprule

    \multirow{2}{*}{Methods} & \multicolumn{3}{c|}{T1c} & \multicolumn{3}{c|}{T2f}  & \multicolumn{3}{c}{Avg} \\
    
    & PSNR $\uparrow$ & SSIM $\uparrow$ & MAE $\downarrow$ & PSNR $\uparrow$ & SSIM $\uparrow$ & MAE $\downarrow$ & PSNR $\uparrow$ & SSIM $\uparrow$ & MAE $\downarrow$\\
    \midrule

    Pix2pix~\cite{isola2017image}	
    & 27.05	& 0.858	& 0.0180	& 24.82	& 0.846& 0.0250& 25.93& 0.852& 0.0215 \\
    CycleGAN~\cite{zhu2017unpaired}	
    & 30.13	& 0.906	& 0.0120	& 26.85	& 0.883	& 0.0188& 28.49	& 0.894 & 0.0154\\
    GcGAN~\cite{fu2019geometry}	
    & 29.98	& 0.917	& 0.0129& 25.98	& 0.872	& 0.0225& 27.98	& 0.894& 0.0177 \\
    CUT~\cite{park2020contrastive}	
    & 26.27	& 0.846	& 0.0181& 23.54	& 0.819	& 0.0278& 24.90& 0.832&0.0229 \\
    RegGAN~\cite{kong2021breaking}	
    & 31.36	& 0.930	&{0.0109}& 29.13& 0.917	& \textbf{0.0135}&30.24& 0.923& {0.0122} \\
    ResViT~\cite{dalmaz2022resvit}	& 31.46	& 0.932	& 0.0131& 28.63	& 0.909	& 0.0166	& 30.04	& 0.920	& 0.0148 \\
    Diffusion~\cite{dhariwal2021diffusion}	& 31.98	& 0.930	& 0.0109	& 29.22	& 0.921	& 0.0155	& 30.60	& 0.925	& 0.0132 \\
    \midrule
    \rowcolor{gray!15}
    Ours	& \textbf{33.08}	& \textbf{0.948}	& \textbf{0.0098}	& \textbf{30.76}	& \textbf{0.934}	& 0.0136	& \textbf{31.92}	& \textbf{0.941}	& \textbf{0.0117} \\
    
    \bottomrule
    \end{tabular}
    }
\end{table*}

\begin{table*}[!t]
    \centering
    \caption{Quantitative comparison on UPenn-GBM dataset} 
    \label{tab:bras_I2I}
    \renewcommand\arraystretch{1.2}
    \resizebox{0.9\textwidth}{!}{
    \begin{tabular}{c | c c c | c c c | c c c}
    \toprule

    \multirow{2}{*}{Methods} & \multicolumn{3}{c|}{T1c} & \multicolumn{3}{c|}{T2f}  & \multicolumn{3}{c}{Avg} \\
    
    & PSNR $\uparrow$ & SSIM $\uparrow$ & MAE $\downarrow$ & PSNR $\uparrow$ & SSIM $\uparrow$ & MAE $\downarrow$ & PSNR $\uparrow$ & SSIM $\uparrow$ & MAE $\downarrow$\\
    \midrule
    
Pix2pix~\cite{isola2017image}	
& 28.93	& 0.925	& 0.0145	& 29.81	& 0.903	& 0.0206	& 29.37	& 0.914	& 0.0175 \\
CycleGAN~\cite{zhu2017unpaired} 
& 30.85	& 0.951	& 0.0122	& 30.25	& 0.914	& 0.0210	& 30.55	& 0.932	& 0.0166 \\
GcGAN~\cite{fu2019geometry}	
& 30.75	& 0.952	& 0.0133	& 30.41	& 0.928	& 0.0209	& 30.58	& 0.940	& 0.0171 \\
CUT~\cite{park2020contrastive}	
& 30.74	& 0.950	& 0.0123	& 31.02	& 0.924	& 0.0187	& 30.88	& 0.937	& 0.0155 \\
RegGAN~\cite{kong2021breaking}	
& 27.71	& 0.916	& 0.0149	& 27.71	& 0.930 	& 0.0163	& 27.71	& 0.923	& 0.0156 \\
Resvit~\cite{dalmaz2022resvit}	
& 27.43	& 0.930	& 0.0138	& 24.58	& 0.905	& 0.0202	& 26.00 	& 0.917	& 0.0170 \\
Diffusion~\cite{dhariwal2021diffusion}	
& 31.84	& 0.962	& 0.0112	& 32.42	& 0.944	& 0.0158	& 32.13	& 0.953	& 0.0135 \\
\midrule
\rowcolor{gray!15} 
Ours & \textbf{33.04} & \textbf{0.967} & \textbf{0.0097} & \textbf{33.65} & \textbf{0.952} & \textbf{0.0138} & \textbf{33.34} & \textbf{0.959} & \textbf{0.0117} \\
    
    \bottomrule
    \end{tabular}
    }
\end{table*}

\begin{figure}[!t]
    \centering
    \includegraphics[width=0.85\linewidth]{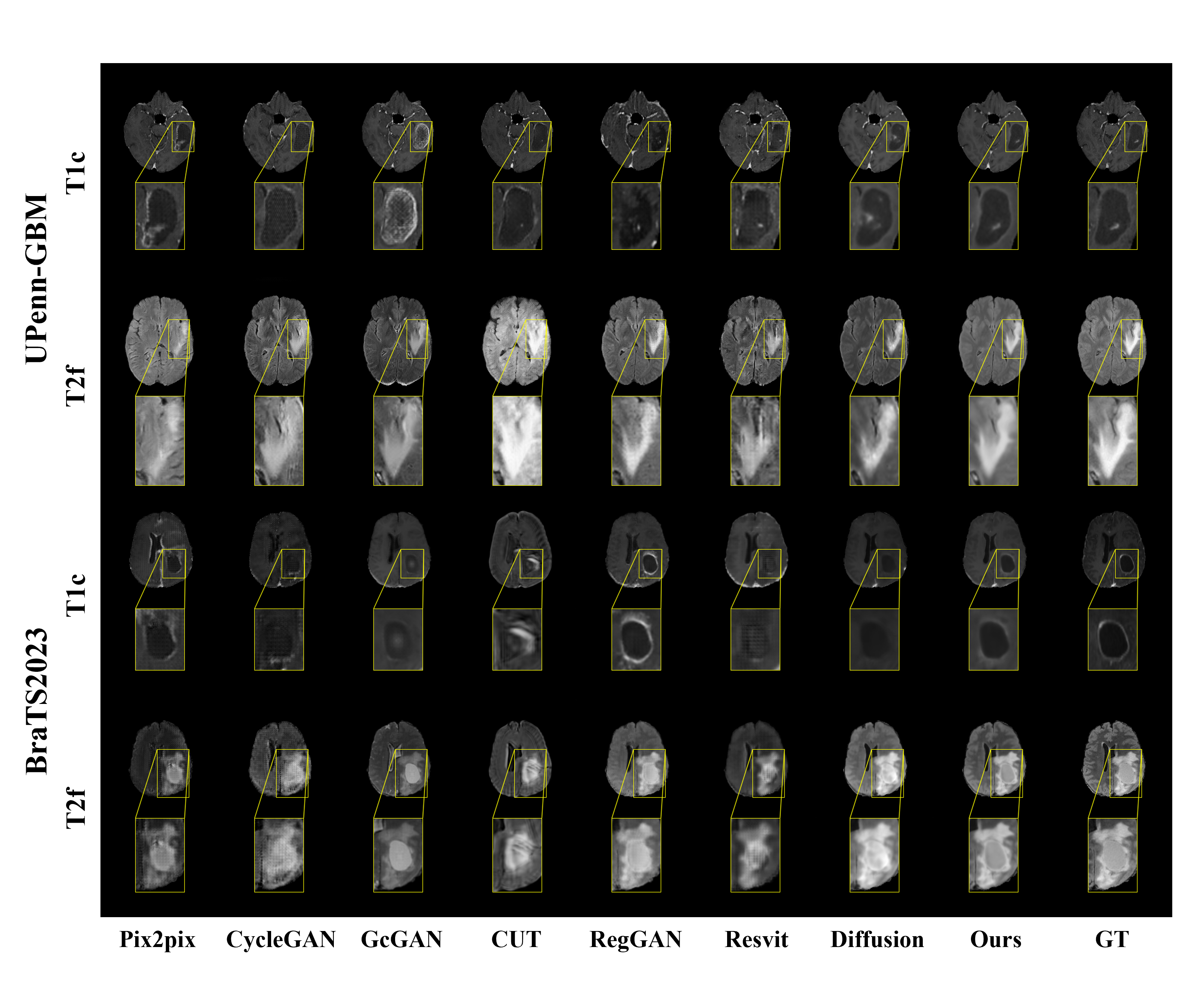}
    \vspace{-2mm}
    \caption{Visual comparisons of proposed CDM and other state-of-the-art methods.}
    \label{fig:show_result}
    \vspace{-3mm}
\end{figure}

\para{BraTS2023} 
Table \ref{tab:bras2023} presents the PSNR, SSIM, and MAE scores for two modalities (T1c and T2f), along with the averaged scores of all methods on BraTS2023. Our CDM achieves the highest PSNR and SSIM scores for T1c and T2f, the lowest MAE score for T1c, and ranks second in MAE for T2f.
More importantly, our method demonstrates superior quantitative performance, averaging 31.92 on PSNR, 0.941 on SSIM, and 0.0117 on MAE, respectively.
Furthermore, we conduct experiments comparing the latest Diffusion method with our method, while our method outperforms the Diffusion method across all metrics.

\para{UPenn-GBM} 
In Table \ref{tab:bras_I2I}, we list PSNR, SSIM, and MAE scores of our network and compared methods for T1c and T2f modalities on the UPenn-GBM dataset, as well as the average metrics.
Among all the comparison methods, the Diffusion has the highest average PSNR and SSIM scores of 32.13 and 0.953, as well as the lowest MAE score of 0.0135. This performance is attributed to the strong representational capabilities of diffusion model.
In comparison, our method has a 3.79\%, 0.63\%, and 13.33\% improvement on PSNR, SSIM, and MAE scores, respectively, achieving state-of-the-art performance.

\begin{table*}[tp]
\small 

\begin{floatrow}
\resizebox{0.48\textwidth}{!}{
\centering
\renewcommand\arraystretch{1.0}
\ttabbox{\caption{Ablation study for different modules on BraTS2023 dataset.}}{%
\label{tab:effectiveness}
\begin{tabular}{c | c c | c c c } 
    \toprule
     Methods & Condition & Decouple & PSNR $\uparrow$ & SSIM $\uparrow$ & MAE $\downarrow$ \\
    \midrule 
    RCG & \Checkmark &  & 15.40 & 0.057 & 0.141 \\
    M1 &   &  &  29.41 & 0.907 & 0.0138 \\
    M2 & \Checkmark &  & 31.73 & 0.935 & 0.0127 \\
    \midrule
    \rowcolor{gray!15} 
    Ours  & \Checkmark & \Checkmark &  \textbf{31.92} & \textbf{0.941} & \textbf{0.0117} \\
    \bottomrule
    \end{tabular}
    }
}

\resizebox{0.48\textwidth}{!}{
\centering
\setlength\tabcolsep{5pt}
\renewcommand\arraystretch{1.2}
\begin{floatrow}
\ttabbox{\caption{Ablation study for different sampling number $N_{sampling}$ on BraTS2023 dataset.}
\label{tab:sample_number}}{%
\begin{tabular}{c | c c c c} 
    \toprule
    $N_{sampling}$ & 10 & 20 & 30 & 40 \\
    \midrule
    PSNR $\uparrow$ & 31.24 & 31.71 & 31.92 & \textbf{31.93}  \\
    SSIM $\uparrow$ &  0.926 & 0.933 & \textbf{0.941} & 0.940 \\
    MAE $\downarrow$ & 0.0123 & 0.0123 & 0.0117 & \textbf{0.0116} \\
    \bottomrule
    \end{tabular}
    }
\end{floatrow}
}
\vspace{-3mm}
\end{floatrow}
\end{table*}

\vspace{-1mm}
\para{Visual Comparisons}
Fig. \ref{fig:show_result} visually compares the synthesis results predicted by our network and state-of-the-art methods
on the BraTS2023 and UPenn-GBM datasets. 
From these visualization results, we can find that our method can more accurately synthesize the brain tumor regions than other methods and maintain the most consistent style with the ground truth.
The reason behind is that our method is capable of learning modality-related information under the guidance of the distributions of T1c and T2f modalities.

\vspace{-3mm}
\subsection{Ablation Study}

\para{The Effectiveness of Each Module}
Table \ref{tab:effectiveness} lists the methods with different modules along with the average PSNR, SSIM, and MAE on T1c and T2f modalities.
As indicated in Table \ref{tab:effectiveness}, RCG has the lowest PSNR and SSIM scores, alongside the highest MAE score. This method employs the encoder pre-trained on natural images, which is not well-suited for medical imaging.
M1 represents our basic method, which only contains a conventional UNet model.
In comparison to M1, M2 integrates the distribution of target modalities (T1c and T2f) as a condition to guide the synthesis process, achieving an improvement of 7.88\%, 3.08\%, and 7.97\% on PSNR, SSIM, and MAE, respectively.
Finally, our method combines both conditional input and the MDN, achieving the state-of-the-art performance of 31.92, 0.941, and 0.0117 across the three metrics.

\para{The Optimal Sampling Number}
To determine the best sampling number $N_{sampling}$, we conduct an experiment in which we increase the sampling number from 10 to 40 with a stride of 10, and evaluate the average PSNR, SSIM, and MAE for T1c and T2f.
As shown in Table \ref{tab:sample_number}, when the sampling number reaches 40, the SSIM score decreases slightly from 0.941 to 0.940, and the improvements of PSNR and MAE are minimal.
Considering the increase in computational cost from $N_{sampling}=30$ to $N_{sampling}=40$, we select $N_{sampling} = 30$ as our default setting.

\begin{figure}[!t]
    \centering
    \includegraphics[width=0.95\linewidth]
    {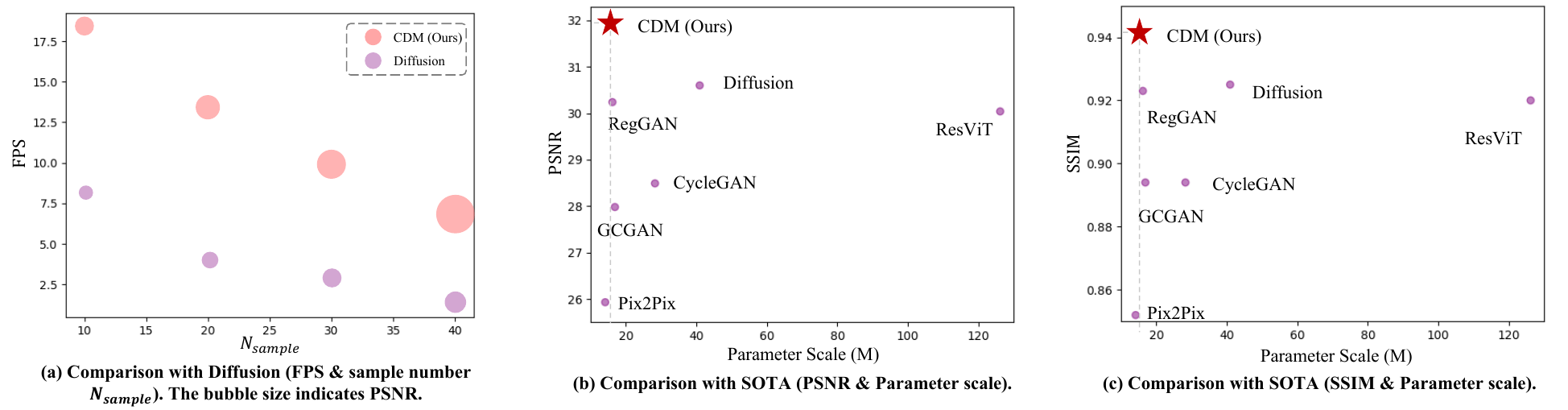}
    \caption{The ablation studies for efficiency and parameter scale.}
    \label{fig:ablation_on_efficiency_parameter}
\end{figure}

\para{The High Efficiency of Our CDM}
Fig. \ref{fig:ablation_on_efficiency_parameter} (a) displays the frames per second (FPS) and PSNR score for both Diffusion and our CDM across different sample number $N_{sample}$ on the BraTS2023 dataset. The larger the bubble size is, the higher the average PSNR for T1c and T2f.
It is observed that our CDM surpasses the Diffusion in terms of both FPS and PSNR at different $N_{sample}$, indicating superior efficiency and synthesis quality.

\para{Comparison on Parameter Scale}
We compare our CDM with state-of-the-art methods in terms of parameter scale, PSNR, and SSIM on the BraTS2023 dataset.
As shown in Fig. \ref{fig:ablation_on_efficiency_parameter} (b) and (c), our method achieves new state-of-the-art results while maintaining a smaller parameter scale.


\vspace{-3mm}
\section{Conclusion}
\vspace{-1mm}
In this paper, we propose a novel paradigm for medical image-to-image translation, named CDM.
The main idea of CDM is to use a modality-specific representation model (MRM) to learn the distribution of target modalities and a modality-decoupled Diffusion network (MDN) to model the distribution from MRM while achieving higher efficiency.
Finally, we propose a cross-conditioned UNet (C-UNet) to receive the source modalities as input and the distribution sampled by MDN as guidance to generate the target modalities.
Extensive experiments on BraTS2023~\cite{menze2014multimodal,bakas2017advancing,kazerooni2023brain} and UPenn-GBM~\cite{bakas2022university} datasets demonstrate the superiority of our proposed CDM.
The ablation studies are conducted to verify the effectiveness of each module and to demonstrate the advantages of our method in terms of efficiency and parameter scale.

\subsubsection{Acknowledgments} 
This work is supported by the Guangzhou-HKUST(GZ) Joint Funding Program (No. 2023A03J0671), the Guangzhou Municipal Science and Technology Project (Grant No. 2023A03J0671), and a General Research Fund of Hong Kong Research Grants Council (project no. 15218521)
\subsubsection{Disclosure of Interests}
The authors declare that they have no competing interests.

\bibliographystyle{splncs04}
\bibliography{refs}

\end{document}